\documentclass[doublecol]{epl2}
\usepackage{amssymb}

\title{A High Robustness and Low Cost Model for Cascading Failures}
\shorttitle{A High Robustness and Low Cost Model for Cascading Failures}
\author{Bing Wang \and Beom Jun Kim}
\shortauthor{B. Wang \and B.J. Kim}

\institute{ Department of Physics, BK21 Physics Research
Division, and Institute of Basic Science, Sungkyunkwan University,
Suwon 440-746, Korea}


\abstract{
We study numerically the cascading failure problem by using
artificially created scale-free networks and the real network
structure of the power grid. The capacity for a vertex is assigned
as a monotonically increasing function of the load (or the
betweenness centrality). Through the use of a simple functional form
with two free parameters, revealed is that it is indeed possible to
make networks more robust while spending less cost. We suggest that
our method to prevent cascade by protecting less vertices is
particularly important for the design of more robust real-world
networks to cascading failures.
}

\begin{document}

\maketitle

The network robustness has been one of the most central topics in
the complex network research~\cite{Internet}. In scale-free
networks, the existence of hub vertices with high degrees has been
shown to yield fragility to intentional attacks, while at the same
time the network becomes robust to random failures due to the
heterogeneous degree distribution~\cite{Albert,Cohen1,Holme,Bing}.
On the other hand, for the description of dynamic processes on top
of networks, it has been suggested that the information flow across
the network is one of the key issues, which can be captured well by
the betweenness centrality or the load~\cite{Loaddis}.

Cascading failures can happen in many infrastructure networks,
including the electrical power grid, Internet, road systems, and so
on. At each vertex of the power grid, the electric power is either
produced or transferred to other vertices, and it is possible that
from some reasons a vertex is overloaded beyond the given capacity,
which is the maximum electric power the vertex can handle. The
breakdown of the heavily loaded single vertex will cause the
redistribution of loads over the remaining vertices, which can
trigger breakdowns of newly overloaded vertices. This process will
go on until all the loads of the remaining vertices are below their
capacities. For some real networks, the breakdown of a single vertex
is sufficient to collapse the entire system, which is exactly what
happened on August 14, 2003 when an initial minor disturbance in
Ohio triggered the largest blackout in the history of United States
in which millions of people suffered without electricity for as long
as 15 hours~\cite{Report}. A number of aspects of cascading failures
in complex networks have been discussed in the
literature~\cite{CasModel,Control1,Control2,alfa,Realdata,Effic1,Vexload,LoadCost,Wu},
including the model for describing cascade
phenomena~\cite{CasModel}, the control and defense strategy against
cascading failures~\cite{Control1,Control2}, the analytical
calculation of capacity parameter~\cite{alfa}, and the modelling of
the real-world data~\cite{Realdata}.
In a recent paper~\cite{Wu}, the
cascade process in scale-free networks 
with community structure has been investigated,
and it has been found that a smaller modularity is
easier to trigger cascade, which implies the importance
of the modularity and community structure in cascading
failures. 

In the research of the cascading failures, the following two issues
are closely related to each other and of significant interests: One
is how to improve the network robustness to cascading failures, and
the other particularly important issue is how to design manmade
networks with a less cost. In most circumstances, a high robustness
and a low cost are difficult to achieve simultaneously. For example,
while a network with more edges are more robust to failures, in
practice, the number of edges is often limited by the cost to
construct them. In brevity, it costs much to build a robust network.
 Very recently, Sch\"{a}fer \emph{et. al.} proposed a new
proactive measure to increase the robustness of heterogeneous loaded
networks to cascades. By defining the load dependent weights, the
network turns to be more homogeneous and the total load is
decreased, which means the investment cost is also
reduced~\cite{LoadCost}. In the present Letter, for simplicity, we
try to find a possible way of protecting networks based on the flow
along shortest-hop path, first proposed by
Motter-Lai~\cite{CasModel}. Through the use of our improved capacity
model, we numerically examine the cascades in scale-free networks
and the electrical power grid network. Since for heterogeneously
loaded networks, overload avalanches can be triggered by the failure
of only one of the most loaded vertices, the following results are
all based on the removal of one vertex with the highest load. Our
results suggest that networks can indeed be made more robust while
spending less cost.

\begin{figure}
{\includegraphics[width=0.9\columnwidth]{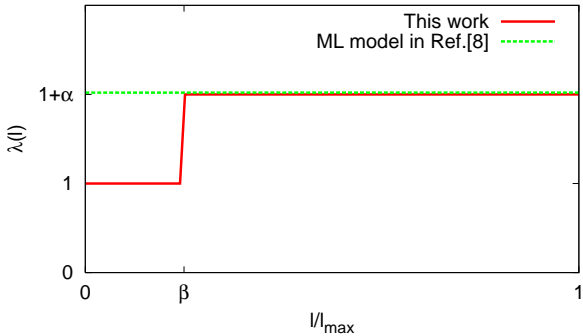}}
\caption{The capacity $c$ is assigned as $c = \lambda(l) l$ with the initial load
$l$. The step function $\lambda(l) = 1 + \alpha \Theta(l/l_{\rm max}
- \beta)$ with two free parameters $\alpha$ and $\beta$ is used in
our model. For comparison, the curve for the Motter-Lai (ML)
capacity model in Ref.~\cite{CasModel}, where $\lambda(l)=$
constant, is also shown. } \label{fig:step}
\end{figure}

We first construct the Barab\'asi-Albert (BA) scale-free
network~\cite{BAmodel} of the size $N=5000$ with the average degree
$\langle k \rangle\approx 4$ to study the cascading failures. The BA
network is characterized by the degree distribution
$p(k)\sim{k^{-\gamma}}$ with the degree exponent $\gamma=3$, and it
has been shown that the load distribution also exhibits the
power-law behavior~\cite{Loaddis}, which means that there exist a
few vertices with very large loads.

The betweenness centrality for each vertex, defined as the total
number of shortest paths passing through it, is used as the measure
of the load and computed by using the efficient
algorithm~\cite{Newman}. The capacity $c_v$ for the vertex $v$ is
assigned as
\begin{equation}
\label{eq:c} c_v=\lambda(l_v) l_v ,
\end{equation}
where $l_v$ is the initial load without failed vertices. Although it
should be possible to find, via a kind of the variational approach,
the optimal functional form of $\lambda(l_v)$ which gives rise to
the lower cost and the higher robustness (see below for the
definitions of the two) we in this work simplify $\lambda(l_v)$ as
shown in Fig.~\ref{fig:step}:
\begin{equation}
\label{eq:step} \lambda(l_v) = 1 + \alpha\Theta(l_v/l_{\rm max} -
\beta),
\end{equation}
where $\Theta(x) = 0 (1)$ for $x < 0 (> 0)$ is the Heaviside step
function, $l_{\rm max} = \max_v l_v$, and we use
$\alpha\in{[0,\infty)}$ and $\beta \in [0,1]$ as two control
parameters in the model. In Ref.~\cite{CasModel} a constant
$\lambda$  has been used (see Fig.~\ref{fig:step} for comparison),
which corresponds to the limiting case of $\beta = 0$ with the
identification $\lambda = 1 + \alpha$ in our model.

At the initial time $t=0$, the vertex with the highest load is
removed from the network, and then new loads for all other vertices
are recomputed.\footnote{In real situations of failures,
the initial breakdown can happen at any vertex in the network.
However, the eventual scale of damages must be greater when
a heavily loaded vertex is broken, and accordingly  we in this work
restrict ourselves to the worst case when the vertex with the highest
load is initially broken.}
We then check the failure condition $c_v < l_v(t)$
for each vertex, and remove all overloaded vertices to get the
network at $t+1$. The above process continues until all existing
vertices fulfill the condition $c_v > l_v(t)$, and the size of the
giant component $N^\prime$ at the final stage is measured. The
relative size of the cascading failures is conveniently captured by
the ratio~\cite{CasModel}
\begin{equation}
\label{eq:g} g=\frac{N^\prime}{N} ,
\end{equation}
which we call the robustness from now on. For networks of
homogeneous load distributions, the cascade does not happen and $g
\approx 1$ has been observed~\cite{CasModel}. Also for networks of
scale-free load distributions, one can have $g \approx 1$ if
randomly chosen vertices, instead of vertices with high loads, are
destroyed at the initial stage~\cite{CasModel}.

In general, one can split, at least conceptually, the total cost for
the networks into two different types: On the one hand, there should
be the initial construction cost to build a network structure, which
may include e.g., the cost for the power transmission lines in power
grids, and the cost proportional to the length of road in road
networks.  Another type of the cost is required to make the {\it
given network} functioning, which can be an increasing function of
the amount of flow and can be named as the running cost. For
example, we need to spend more to have bigger memory sizes and
faster network card and so on for the computer server which delivers
more data packets. In the present Letter, we assume that the network
structure is given, (accordingly the construction cost is fixed),
and focus only on the running cost which should be spent in addition
to the initial construction cost.

Without consideration of the cost to protect vertices, the cascading
failure can be made never to happen by assigning extremely high
values to capacities. However, in practice, the capacity is severely
limited by cost.
We expect the cost to protect the vertex $v$ should
be an increasing function of $c_v$, and for convenience define the
cost $e$ as
\begin{equation}
\label{eq:e} e = \left[\sum_{v=1}^N \bigl(\lambda(l_v) -1 \bigr)
\right]/N.
\end{equation}
It is to be noted that for a given value of $\alpha$, the original
Motter-Lai (ML)  capacity model in Ref.~\cite{CasModel} has always a
higher value of the cost than our model (see Fig.~\ref{fig:step}).
Although $e = 0$ at $\beta = 1$, it should not be interpreted as a
costfree situation; we have defined $e$ only as a relative measure
in comparison to the case of $\lambda(l) = 1$ for all vertices. For
a given network structure, the key quantities to be measured are
$g(\alpha,\beta)$ and $e(\alpha,\beta)$, and we aim to increase $g$
and decrease $e$, which will eventually provide us a way to achieve
the high robustness and the low cost at the same time.

\begin{figure}
\begin{center}
{\includegraphics[width=0.9\columnwidth]{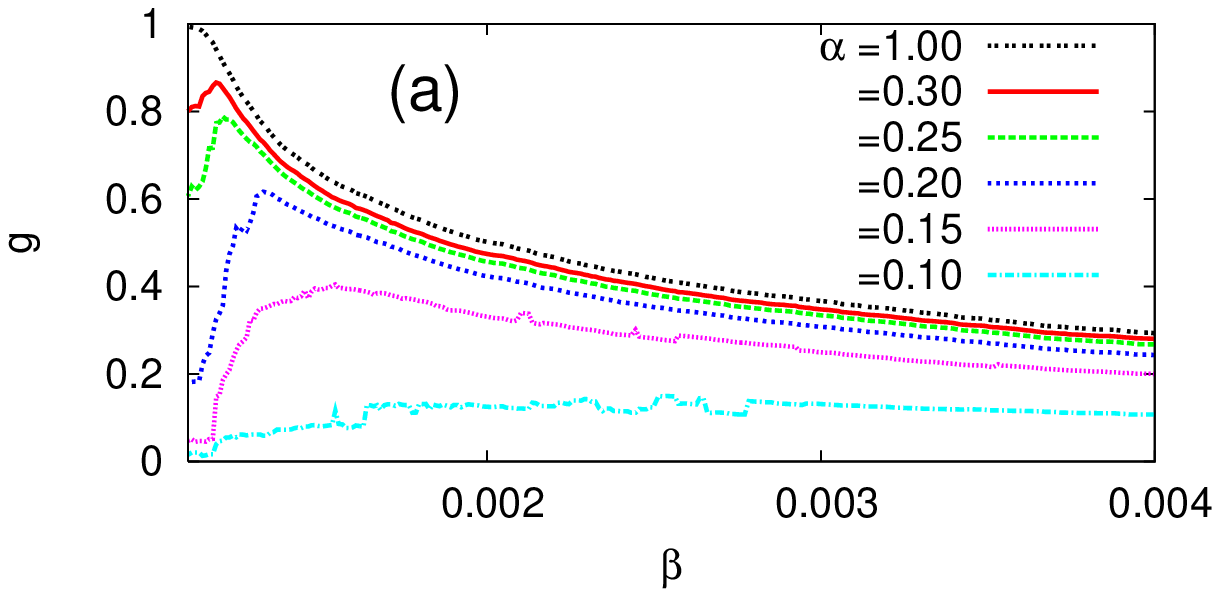}}
{\includegraphics[width=0.9\columnwidth]{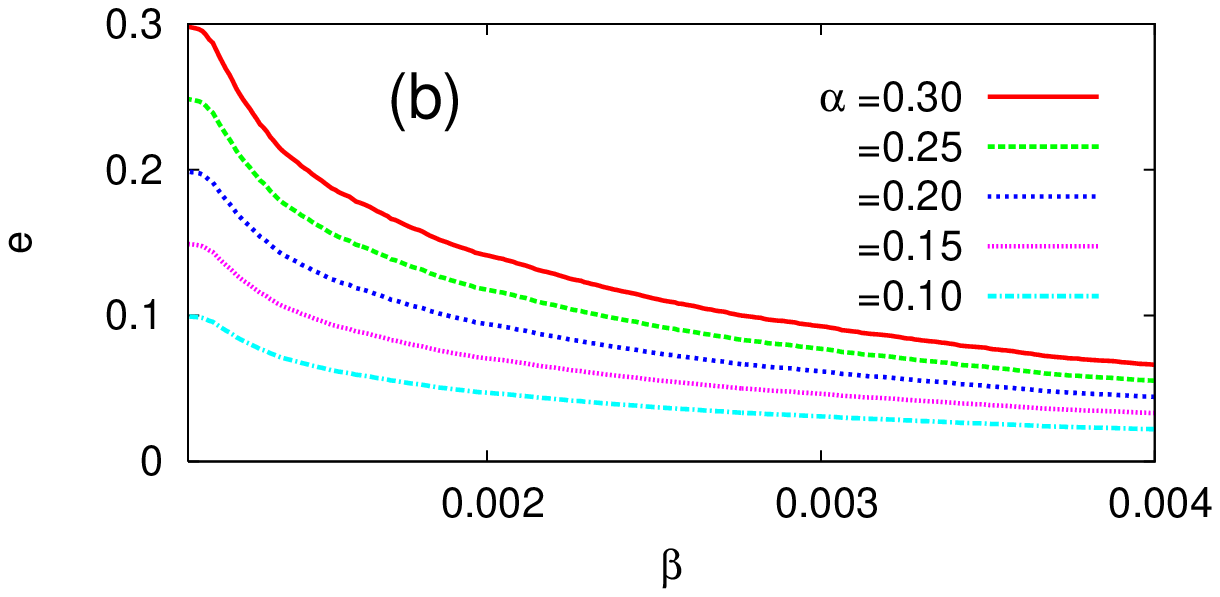}}
{\includegraphics[width=0.9\columnwidth]{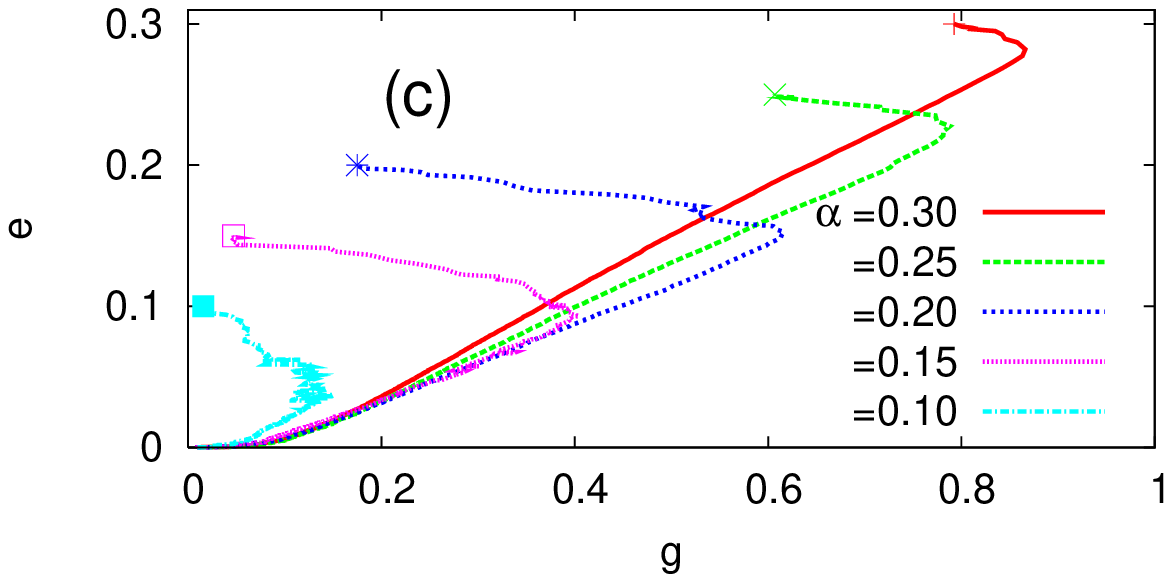}} \caption{
Cascading failures in the BA network of the size $N=5000$ and the
average degree $\langle k \rangle \approx 4$, triggered by the
removal of a single vertex with the highest load. The robustness $g$
and the cost $e$ in Eqs.~(\ref{eq:g}) and (\ref{eq:e}) are shown in
(a) and (b), respectively, as functions of $\beta$ at various
$\alpha$ values [see Fig.~\ref{fig:step} for $\alpha$ and $\beta$,
the two parameters in the function $\lambda(l)$ in
Eq.~(\ref{eq:step})]. (c) The relation between $e$ and $g$ at
different $\alpha$'s.  Compared with the ML model in
Ref.~\cite{CasModel}, it is clearly shown that the network can be
made more robust but with less cost. } \label{fig:BA}
\end{center}
\end{figure}

In Fig.~\ref{fig:BA}(a), we report the robustness $g$ for the BA
network of the size $N=5000$ with the average degree $\langle k
\rangle \approx 4$  as a function of $\beta$ at $\alpha = 0.10$,
0.15, 0.20, 0.25, 0.30, and 1.0 (from bottom to top). As $\beta$
increases further beyond the region in Fig.~\ref{fig:BA}(a), the
robustness $g$ is found to decrease toward zero (not shown here),
which is as expected since the larger $\beta$ makes vertices with
larger loads less protected (see Fig.~\ref{fig:step}). We also skip
in Fig.~\ref{fig:BA} small values of $\beta$ below approximately
0.001: If $\beta < l_{\rm min}/l_{\rm max}$, with the minimum load
$l_{\rm min}$, all vertices are given $\lambda(l) = 1+\alpha$,
equivalent to the ML model corresponding to $\beta=0$. It is shown
in Fig.~\ref{fig:BA}(a) that for $\alpha\lesssim{0.30}$, $g$ first
increases and then decreases as $\beta$ is increased, exhibiting a
well-developed maximum $g_{\rm max}$ at $\beta = \beta^*$. This is a
particularly interesting observation since {\em the network becomes
more robust (larger $g$) by protecting less vertices (larger
$\beta$)}.
In more detail, the curve for $\alpha = 0.20$ in
Fig.~\ref{fig:BA}(a) shows the maximum $g_{\rm max} \approx 0.62$
(at $\beta^* \approx 0.00133$), which is about 3.5 times bigger than
$g\approx 0.175$ (at $\beta = 0$). In other words, the network can
be made much more robust by assigning smaller capacities to vertices
with less loads. For larger values of $\alpha$, on the other hand,
it is found that $g_{\rm max}$ occurs at $\beta = 0$, which
indicates that the above finding, i.e., possibility of making
network more robust by protecting less vertices, does not hold, as
exemplified by the curve for $\alpha = 1$ in Fig.~\ref{fig:BA}(a).

The above observation is closely related with Ref.~\cite{Control1},
where it has been found that in order to reduce the size of cascades
(or to have a larger $g$), some of less loaded vertices should be
removed just after the initial attack. In reality, however, we
believe that the direct application of this strategy of intentional
breakdowns is not easy, for cascading failures usually propagate
across the whole network very soon just after the initial breakdown.
In contrast, we propose in this work a way to make the network
better prepared to breakdowns, by protecting less vertices.

In order to look at the cost benefit of protecting less vertices in
a more careful way, we plot in Fig.~\ref{fig:BA}(b) the cost $e$ in
Eq.~(\ref{eq:e}) versus $\beta$ at various values of $\alpha$. As is
expected from Fig.~\ref{fig:step}, the cost $e$ is shown to be a
monotonically decreasing (increasing) function of $\beta$ ($\alpha$)
at fixed $\alpha$ ($\beta$). Take again the case with $\alpha =
0.20$ as an example with $e(\beta^*) \approx 0.153$ and $e(\beta=0)
= 0.2$: It is then concluded that for $\alpha = 0.2$ one can make
the network 3.5 ($\approx 0.62/0.175$) times more robust while
spending only 76.5\% ($\approx 0.153/0.2)$ of the original cost.

In Fig.~\ref{fig:BA}(c), we use the same data as in
Fig.~\ref{fig:BA}(a) and (b), and show the relation between the
robustness and the cost for $\alpha=0.10, \cdots, 0.30$ from bottom
to top. For comparison, the values ($g$,$e$) for $\beta = 0$,
corresponding to the ML model, are also displayed as symbols at the
end of curves. It is clearly shown that for a given $\alpha$, one
can achieve the higher robustness and the lower cost by tuning
$\beta$ toward the right-most point on each curve. We can also use
Fig.~\ref{fig:BA}(c) to choose  the most efficient way to get a
given robustness $g$: For example, suppose that $g = 0.6$ is the
required robustness. The vertical line for $g=0.6$ crosses several
different curves, and one can choose the crossing point which has
the lowest cost.

\begin{figure}
\begin{center}
{\includegraphics[width=0.9\columnwidth]{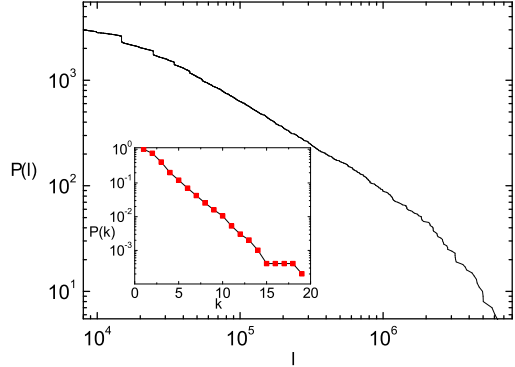}} \caption{The
cumulative load distribution of power grid network $P(l)$ in log-log
scale. The inset shows the cumulative degree distribution $P(k)$ of
the power grid in linear-log scale. } \label{fig:powerdist}
\end{center}
\end{figure}
\begin{figure}
\begin{center}
{\includegraphics[width=0.9\columnwidth]{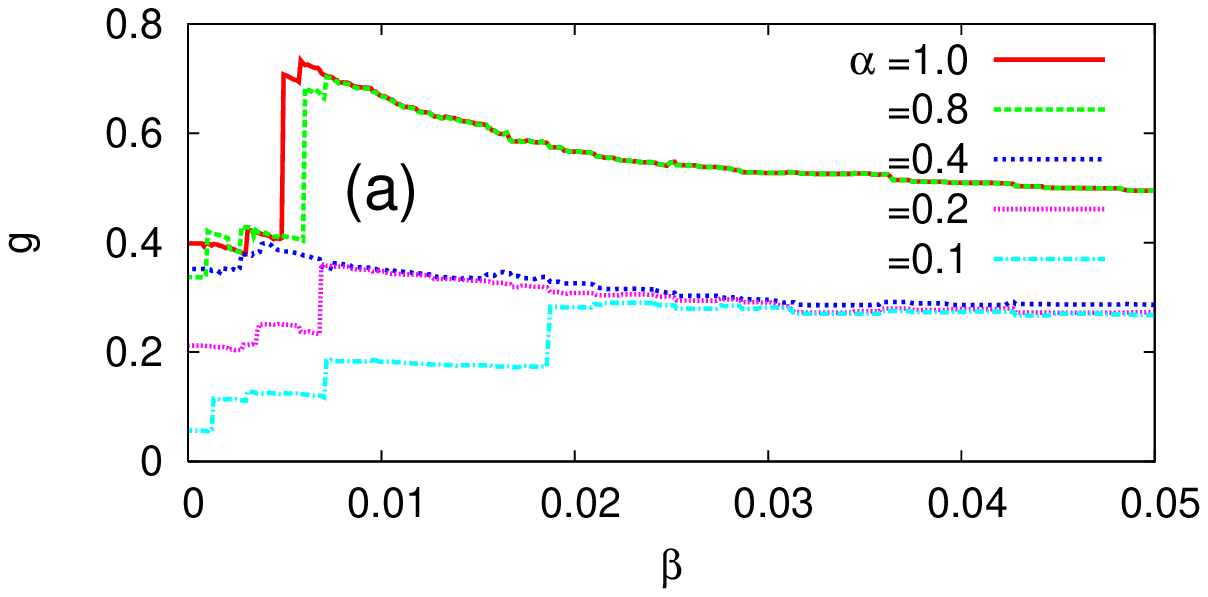}}
{\includegraphics[width=0.9\columnwidth]{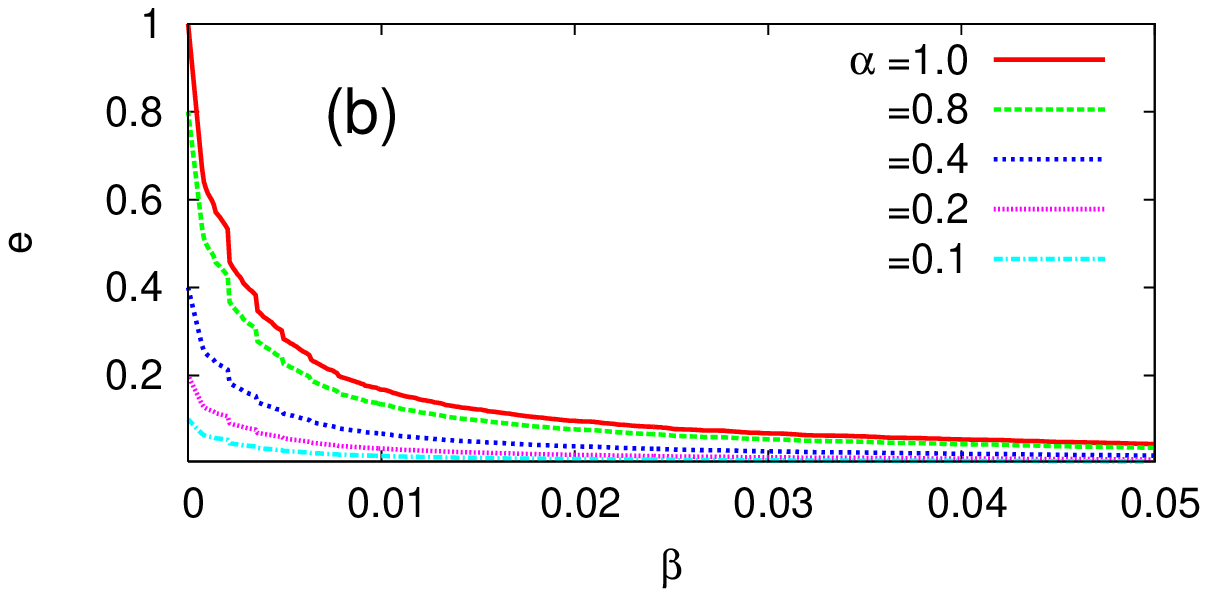}}
{\includegraphics[width=0.9\columnwidth]{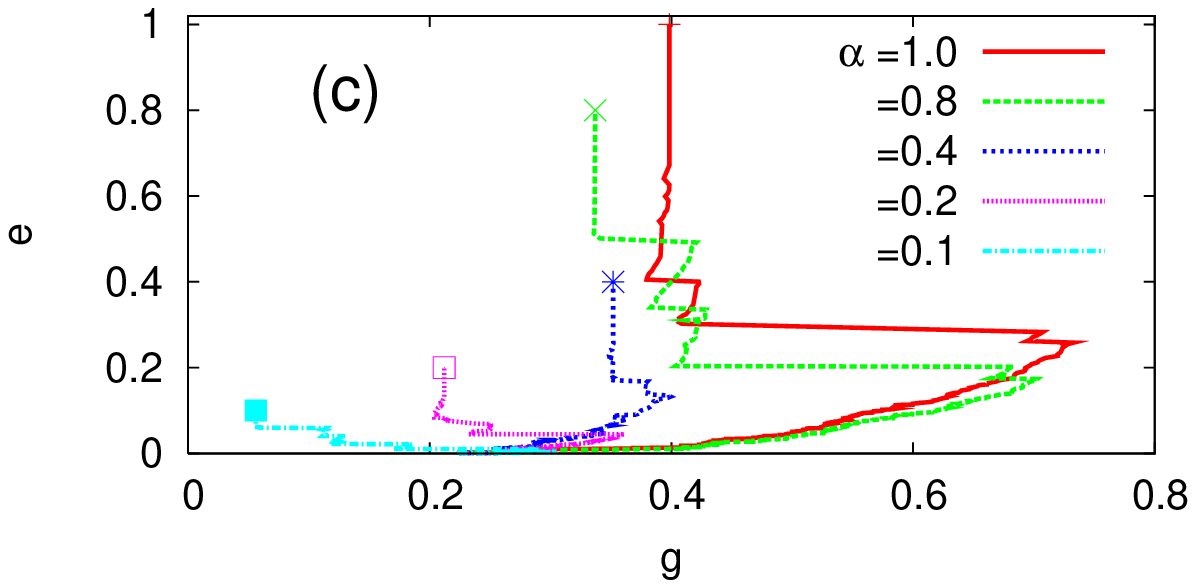}} \caption{
Cascading failures in the electrical power grid of the size
$N=4941$. (Compare with Fig.~\ref{fig:BA} for the corresponding
plots for the BA network.) The robustness $g$  and the cost $e$
versus $\beta$ at various $\alpha$ values are shown in (a) and (b),
respectively, while (c) is for the relation between $e$ and $g$.
Again, it is shown that one can achieve the higher robustness and
the less cost simultaneously, by choosing the right-most point in
(c). } \label{fig:power}
\end{center}
\end{figure}

We next study the cascading failures in the real network structure
of the North American power grid of the size
$N=4941$~\cite{Powerdata}. Although the electrical power grid
network is a very homogeneous network in terms of the degree
distribution, the load distribution, in a sharp contrast, shows a
strong heterogeneity as shown in Fig.~\ref{fig:powerdist}. In other
words, the degree distribution is more like an exponential one,
while the load distribution is similar to the power-law form. The
broad load distribution  can be one of the reasons of the fragility
of the power grid to cascading failures~\cite{CasModel}.

We then apply, the same method as we used above, to the power grid,
and obtain $g$ and $e$ as functions of $\beta$ for given values of
$\alpha$. Figure~\ref{fig:power} for the cascading failures of the
power grid is in parallel to Fig.~\ref{fig:BA} for the BA network:
Fig.~\ref{fig:power}(a) for $g$ versus $\beta$, (b) for $e$ versus
$\beta$, and (c) for $e$ versus $g$. There are some quantitative
differences between curves for the power grid  and the BA network.
However, qualitatively speaking, both networks are shown to exhibit
the following common features: (i) For a given $\alpha$, the
robustness has a maximum $g_{\rm max}$ at $\beta=\beta^*$, (ii) $e$
is a monotonically decreasing  function of $\beta$  at a given
$\alpha$, and (iii) there exists a lob-like structure in the $g$-$e$
plane, which indicates that one can make the network exhibit a
higher robustness and a lower cost at the same time than the
corresponding values for the ML model. It is worth mentioning that
the power grid in Fig.~\ref{fig:power} can be made to show the
higher $g$ and the lower $e$ than the ML model in a broader region
of $\alpha$: Even at $\alpha = 1$, the power grid can have much
better robustness and much less cost in comparison to the ML model.
Specifically, at $\alpha = 1.0$ the ML model has $g \approx 0.40$
and $e = 1.0$ while our model can yield $g \approx 0.73$ and $ e
\approx 0.26$ (at $\beta \approx 0.00583$) [see
Fig.~\ref{fig:power}(c)], which occurs when only 26\% of vertices
are given the higher capacity $\lambda(l) = 2$, and the other
remaining 74\% of vertices have the lower capacity $\lambda(l) = 1$.
In other words, by assigning  lower capacities to 74\% of vertices,
the network becomes much more robust.

\begin{figure}
\begin{center}
{\includegraphics[width=0.9\columnwidth]{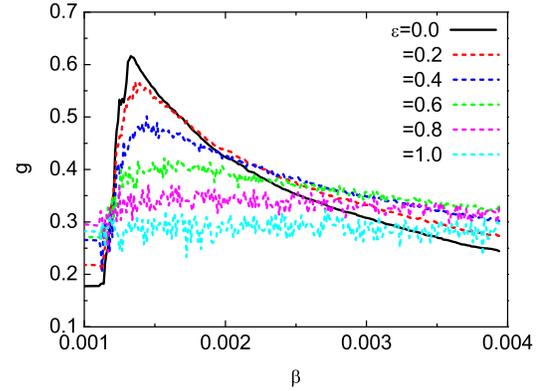}}
 \caption{
Cascading failures in the BA network of the size $N=5000$ and the
average degree $\langle k \rangle\approx{4}$, triggered by the
removal of a single vertex with the highest load. Each vertex's
capacity is disturbed with probability $\varepsilon$ for
$\alpha=0.2$. The data are averaged over 20 runs. }
\label{fig:noise}
\end{center}
\end{figure}

In reality, it is also interesting to observe the effect of noise on
the dynamical process. In Ref.~\cite{Fogedby}, 
when noise is introduced into the nonlinear dynamical system, 
it has been shown that noise changes the
singularity at a special time to a statistical time distribution and
shows various interesting behaviors. 
In the present work, we are
interested in how the presence of noise influences the final
cascading failure behavior within our scheme. Here, we introduce
effects of noise as an erroneous assignment of the capacity
function. In detail, at a given error probability $\varepsilon$,
the vertex $v$ is assigned the capacity $c_v'$ instead of its
correct $c_v$:
\begin{equation}
c_v'=c_v ( 1 + r),
\end{equation}
where $r$ is the uniform random variable with zero mean
($r\in{[-1,1]})$. We believe that this erroneous behavior is
plausible in reality, since the perfect knowledge for the true
value of the load for each vertex may not be available, which may cause an
erroneous assignment of the capacity  on a vertex. In the limiting
case of $\varepsilon = 0$, we recover our error-free results
presented above. In Fig.~\ref{fig:noise}, we report the results at
$\alpha=0.2$ for the robustness $g$ for the BA network as a function of
$\beta$ for different error probability $\varepsilon$ [see
Fig.\ref{fig:BA}(a) for comparison]. It is seen that for small
$\varepsilon$, the overall behavior is qualitatively the same as in
Fig.~\ref{fig:BA}(a), i.e., the existence of a well-developed
robustness peak and gradual decrease as $\beta$ is increased. The
peak height of the robustness is found to decrease as $\varepsilon$
is increased, indicating the negative effect of the noise. An
interesting observation in Fig.~\ref{fig:noise} is that as
$\varepsilon$ becomes larger there exits a region of $\beta$ in
which the robustness is actually higher than the error-free case of
$\varepsilon = 0$.

In summary, we have suggested a new capacity model to cascading
failures, by improving the existing ML capacity model in
Ref.~\cite{CasModel}. The main idea in our model is the same as in
existing studies: In a highly heterogeneous network with a broad
load distribution, vertices with large loads should be more
protected by assigning large capacities. Different from other
studies in which the capacity is assigned in proportion to the load,
i.e., $c = \lambda  l$, we generalize the model so that the
proportionality constant $\lambda$ is now changed to an increasing
function $\lambda(l)$ of $l$. In more detail, we use the Heaviside
step function for $\lambda(l)$ characterized by two parameters, the
step height $\alpha$, and the step position $\beta$. By applying
this capacity model to the artificial BA network as well as the real
network of the power grid, we have clearly shown that it is indeed
possible to make the network more robust, while at the same time the
cost to assign capacities is drastically reduced. We believe that
our suggested model to assign capacities to vertices should be
practically useful in designing infrastructure networks in an
economic point of view. 
As a final remark, it needs to be
pointed out that the model proposed in this work should be
considered as only the first step to find the optimal functional
form $\lambda(l)$ of the capacity as a function of the load. As a
future work, we are planning to apply a sort of variational method
to find the optimal functional form of  $\lambda(l)$.

B.J.K. was supported by grant No. R01-2005-000-10199-0 from the
Basic Research Program of the Korea Science and Engineering
Foundation.

\end{document}